\documentclass[a4paper,pre,superscriptaddress,onecolumn]{revtex4}
\usepackage[british]{babel}
\usepackage[latin1]{inputenc}
\usepackage{graphicx}
\usepackage{amsmath}

\begin{document}

\title{Dansgaard-Oeschger events:  tipping points in the climate system}

\author{A.~A.~Cimatoribus}
\email{cimatori@knmi.nl}
\affiliation{Royal Netherlands Meteorological Institute, Wilhelminalaan 10, 3732GK De Bilt, Netherlands}
\author{S.~S.~Drijfhout}
\affiliation{Royal Netherlands Meteorological Institute, Wilhelminalaan 10, 3732GK De Bilt, Netherlands}
\author{V.~Livina}
\affiliation{School of Environmental Sciences, University of East Anglia, Norwich NR4 7TJ, UK}
\author{G.~van der Schrier}
\affiliation{Royal Netherlands Meteorological Institute, Wilhelminalaan 10, 3732GK De Bilt, Netherlands}

\begin{abstract}
Dansgaard--Oeschger events are a prominent mode of variability in the records of the last glacial cycle.
Various prototype models have been proposed to explain these rapid climate fluctuations, and no agreement has emerged on which may be the more correct for describing the paleoclimatic signal.
In this work, we assess the bimodality of the system reconstructing the topology of the multi--dimensional attractor over which the climate system evolves.
We use high--resolution ice core isotope data to investigate the statistical properties of the climate fluctuations in the period before the onset of the abrupt change.
We show that Dansgaard--Oeschger events have weak early warning signals if the ensemble of events is considered.
We find that the statistics are consistent with the switches between two different climate equilibrium states in response to a changing external forcing (e.g.~solar, ice sheets\ldots), either forcing directly the transition or pacing it through stochastic resonance.
These findings are most consistent with a model that associates Dansgaard--Oeschger with changing boundary conditions, and with the presence of a bifurcation point.
\end{abstract}

\maketitle

\section{Introduction}
A dominant mode of temperature variability over the last sixty thousand years is connected with Dansgaard--Oeschger events (DOs) \citep{Dansgaard1993,EPICA2004}.
These are fast warming episodes (in the North Atlantic region $5\div10\mathrm{^\circ C}$ in a few decades), followed by a gradual cooling that lasts from hundreds to thousands of years, often with a final jump back to stadial condition.
The spectral power of their time series shows a peak at approximately 1,500 years and integer multiples of this value, suggesting the presence of such a periodicity in the glacial climate \citep{Alley2001,Ganopolski2002}.
The relation between DOs and large changes in the Atlantic meridional overturning circulation is generally considered well established, despite the limitations of the paleoclimatic records \citep{Keigwin1994,Sarnthein1994,Sakai1999,Ganopolski2001,Timmermann2003}.
Different low--dimensional models have been proposed to explain these rapid climate fluctuations \citep{Sakai1999,Ganopolski2001,Timmermann2003,Ditlevsen2010}, linking DOs to ocean and ice sheet dynamics.
No clear evidence in favour of one of them has emerged from observational data.
Here, we use high--resolution ice core isotope data \citep{NGRIP2004,Svensson2008} to investigate the statistical properties of the climate fluctuations \citep{Held2004,Livina2007,Scheffer2009} in the period before the onset of the abrupt change.
We analyse $\delta^{18}O$ isotope data from the NGRIP ice core \citet{NGRIP2004,Svensson2008}.
The data spans the time interval from 59,420 to 14,777 years before 2,000~A.D. (years b2k) with an average resolution of 2.7 years (more than 80\% of the data has a resolution better than 3.5 years).
$\delta^{18}O$ can be interpreted as a proxy for atmospheric temperature over Greenland.
We use the dating for the onset of the interstadials given in \citet{Svensson2008}.
The dataset then spans the events numbered 2 to 16.

In the first part of the paper, we discuss the bimodality of the time series, already demonstrated in other works~\citep{Livina2011,Livina2010,Wunsch2003}.
In particular, we establish that bimodality is a robust feature of the climate system that produced it, and not an artifact of the projection of complex dynamics onto a scalar quantity.
This is done using a phase embedding technique well known in the non--linear dynamics community, but not often used in climate sciences.

In the second part of the paper, we show that the statistical properties of the paleoclimatic time series can be used to distinguish between the various models that have been proposed to explain DOs.
We suggest that considering the properties of an ensemble of events may uncover signals otherwise hidden, and we show that this seems to be the case for the DOs in the time series considered here.
Within the limitations of the available data and of the techniques used, we find that the statistics are most compatible with a system that switches between two different climate equilibrium states in response to a changing external forcing \citep{Paillard1994}, or with stochastic resonance \citep{Ganopolski2002}.
In both cases, the external forcing controls the distance of the system from a bifurcation point. 
It must be clarified here that we use ``external forcing'' in a purely mathematical sense: the presence of an external forcing means that the climate system can be described by a non--autonomous system of equations, i.e. the evolution of the system explicitely depends on time, as a forcing term is present \citep[see e.g.][]{Arfken2005}.
No assumption on the nature of this forcing is made.
Other hypotheses \citep[noise-induced transitions and autonomous oscillations, suggested by][]{Timmermann2003,Sakai1999,Ditlevsen2010} are less compatible with the data.

\section{Bimodality}
\label{part:Bimodality}

The bimodality hypothesis for this climate proxy has been tested in previous works~\citep{Livina2011,Livina2010,Wunsch2003}, but these make the strong assumption that the bimodality of the recorded time series can be directly ascribed to the climate system that produced it.
This is not necessarily the case if the evolution of a system with many degrees of freedom, as the real climate system, is projected on a scalar record.
Exploiting Takens' embedding theorem \citep{Broer2011} for reconstructing phase space (the space that includes all the states of a system), we assess bimodality without making any assumption on the underlying model beforehand.
This approach has the advantage of avoiding artifacts that may arise from the projection of a higher dimensional system onto a scalar time series.

\subsection{Phase space reconstruction technique}

\begin{figure}[t]
  \begin{center}
    \setlength\fboxsep{0pt}
    \setlength\fboxrule{0.5pt}
 
    \fbox{\includegraphics[trim=10mm 5mm 15mm 15mm,clip,width=0.49\textwidth]{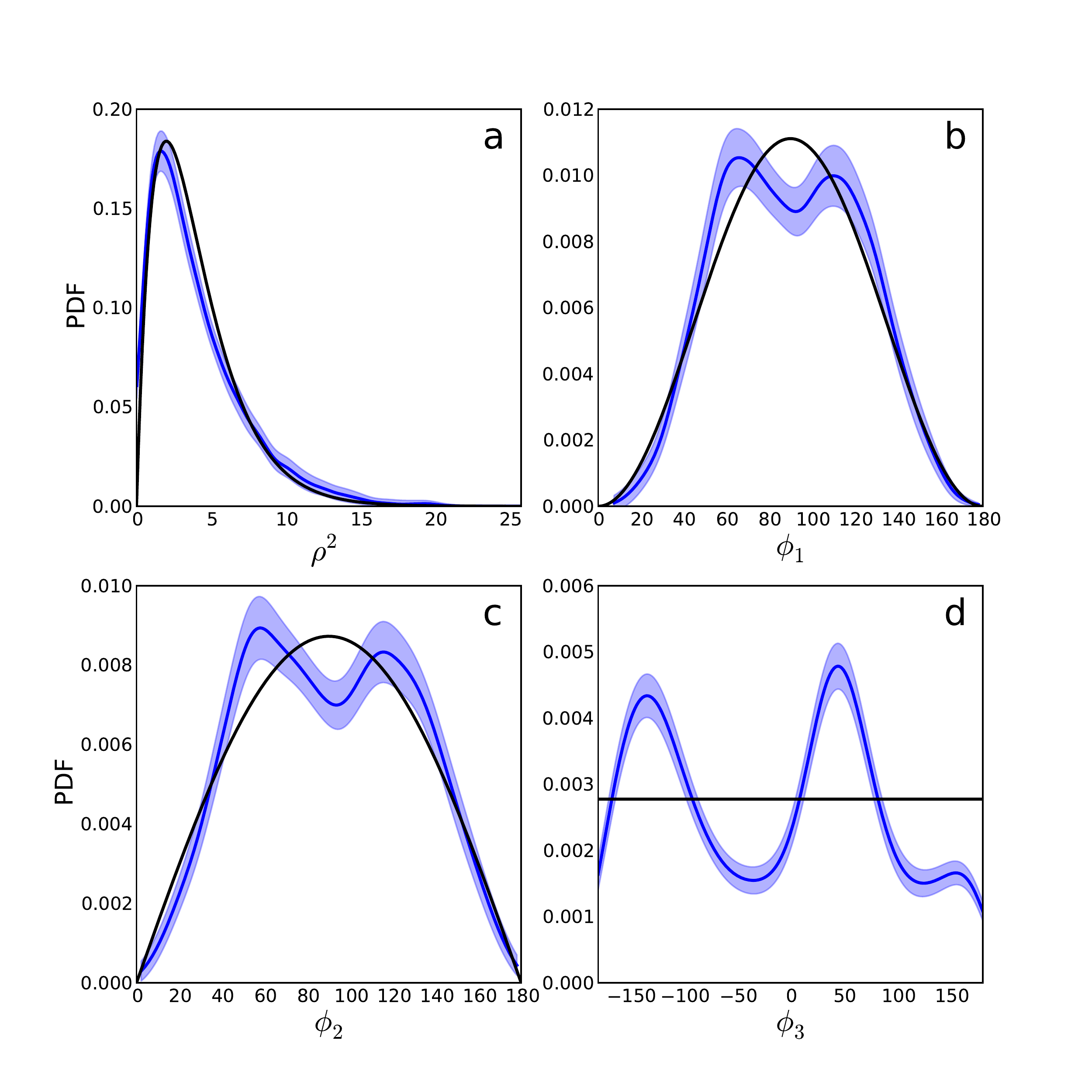}}
    \fbox{\includegraphics[trim=10mm 5mm 15mm 15mm,clip,width=0.49\textwidth]{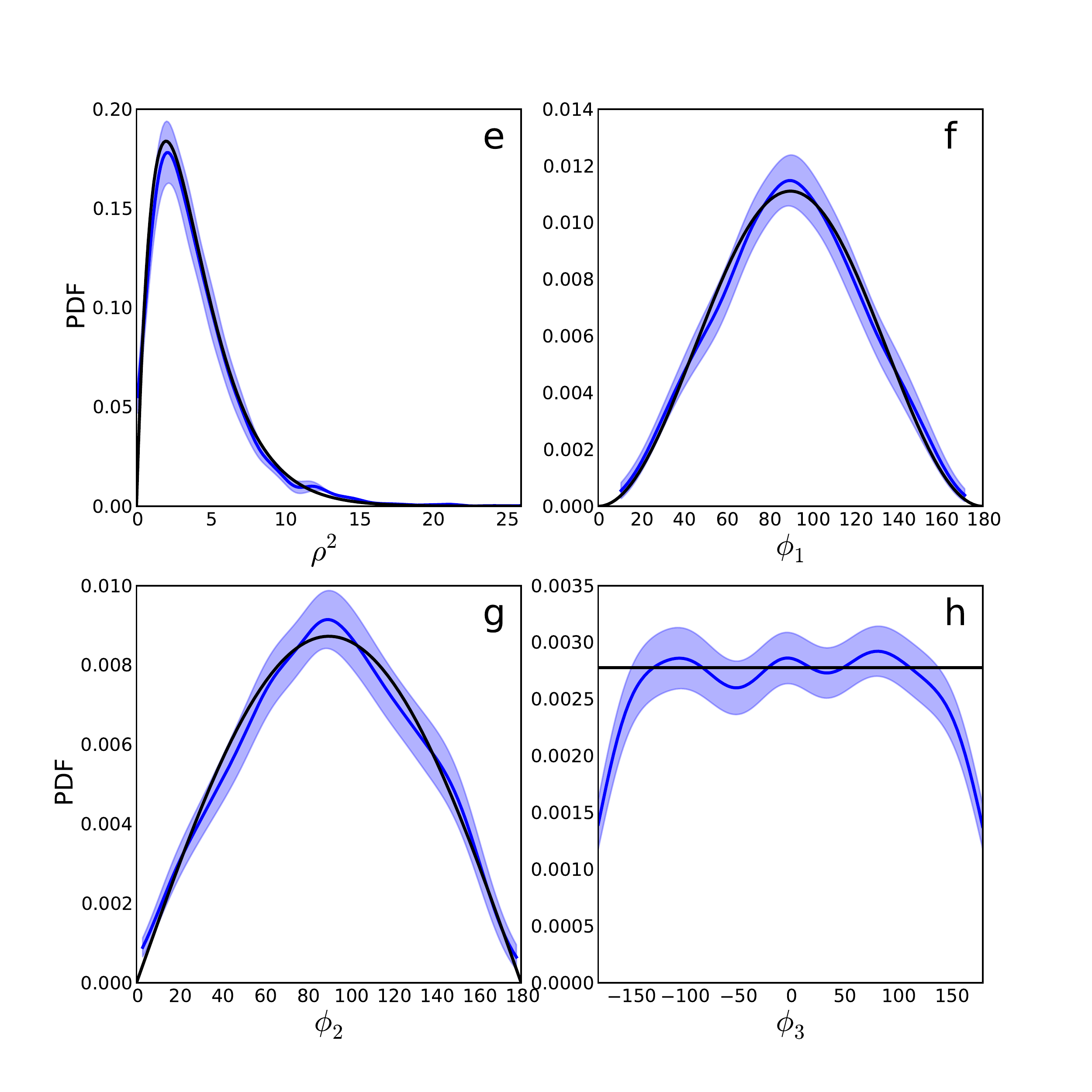}}
    \caption{Phase space reconstruction of linearly detrended data, before and after year 22,000 b2k.
      Blue lines denote kernel estimators of PDFs, with error margins shaded (see text).       
      The left (a, b, c and d) and right panels (e, f, g, h) are, respectively, the phase space reconstruction for the portion of data before and after year 22,000 b2k.
      PDFs for the four coordinates of the phase space reconstruction are plotted.
      The origin of the space is the barycentre of the data, 4--dimensional polar coordinates are used.
      a and e are the PDF for the square of the radial coordinate.
      b--c and f--g are the PDFs for the two ``latitude--like'' coordinates.
      d and e are the PDF for the ``longitude--like'' coordinate.
      The black lines are the theoretical PDFs for a unimodal multinormal sample: in the polar reference system, a $\chi^2$ distribution, with number of degrees of freedom equal to the number of dimensions, for the square of the radius ($\rho^2$).
      For the angular coordinates, a unimodal multinormal PDF would be proportional to $\mathrm{sin}^2( \phi_1)$ and $\mathrm{sin} (\phi_2)$ respectively, while it would be uniform in $\phi_3$.
      \label{fig:Bimodality}
    }
  \end{center}
\end{figure}

The technique that we apply is outlined in~\citet{Abarbanel1996}, and has some points of contact with the Singular Spectrum Analysis described e.g.~in \cite{Ghil2002}.
First, the irregular time series is linearly interpolated to a regular time step, 4 years in the results shown here.
Data after the last DO is discarded for producing the phase space reconstruction.
After detrending, an optimal time lag has to be empirically chosen, guided by the first minimum of the average mutual information function (20 years in our case).
With copies of the regular time series lagged by multiples of the optimal time lag as coordinates, a phase space reconstruction is obtained.
The dimension of the embedding space is, in general, greater than or equal to the one of the physical system which produced the signal, and the dynamics of the system are thus completely unfolded.
The number of dimensions is chosen as the minimum one that brings the fraction of empirical global false neighbours to the background value~\citep{Kennel1992}.
In our reconstructions, four dimensions are used.
We define as global false neighbours in dimension $n$ those couples of points whose distance increases more than 15 times, or more than two times the standard deviation of the data, when the embedding dimension is increased by one.
The Scientific Tools for Python (SciPy) implementation of the kd--tree algorithm is used for computing distances in the \emph{n}--dimensional space.
The 4--dimensional phase space reconstruction is presented in Figure~\ref{fig:Bimodality}, performed separately for the data before and after year 22,000 b2k.
The results are converted to a polar coordinate system and the anomaly from the average is considered, following~\citet{Stephenson2004}.
Subsequently, the PDFs for the coordinates distributions are computed through a gaussian kernel estimation (blue line in Fig.~\ref{fig:Bimodality}).
The error of the PDFs is defined as the standard deviation of an ensemble of 50,000 synthetic datasets (shaded region in the figure), obtained by bootstrapping \citep[see e.g.][]{Efron1982}.
The theoretical distributions (black lines) refer to a multidimensional normal distribution (the simplest null hypothesis, following \citet{Stephenson2004}).

\subsection{Results: bimodality and bifurcation}

\begin{figure}[t]
  \begin{center}
    \includegraphics[width=\linewidth]{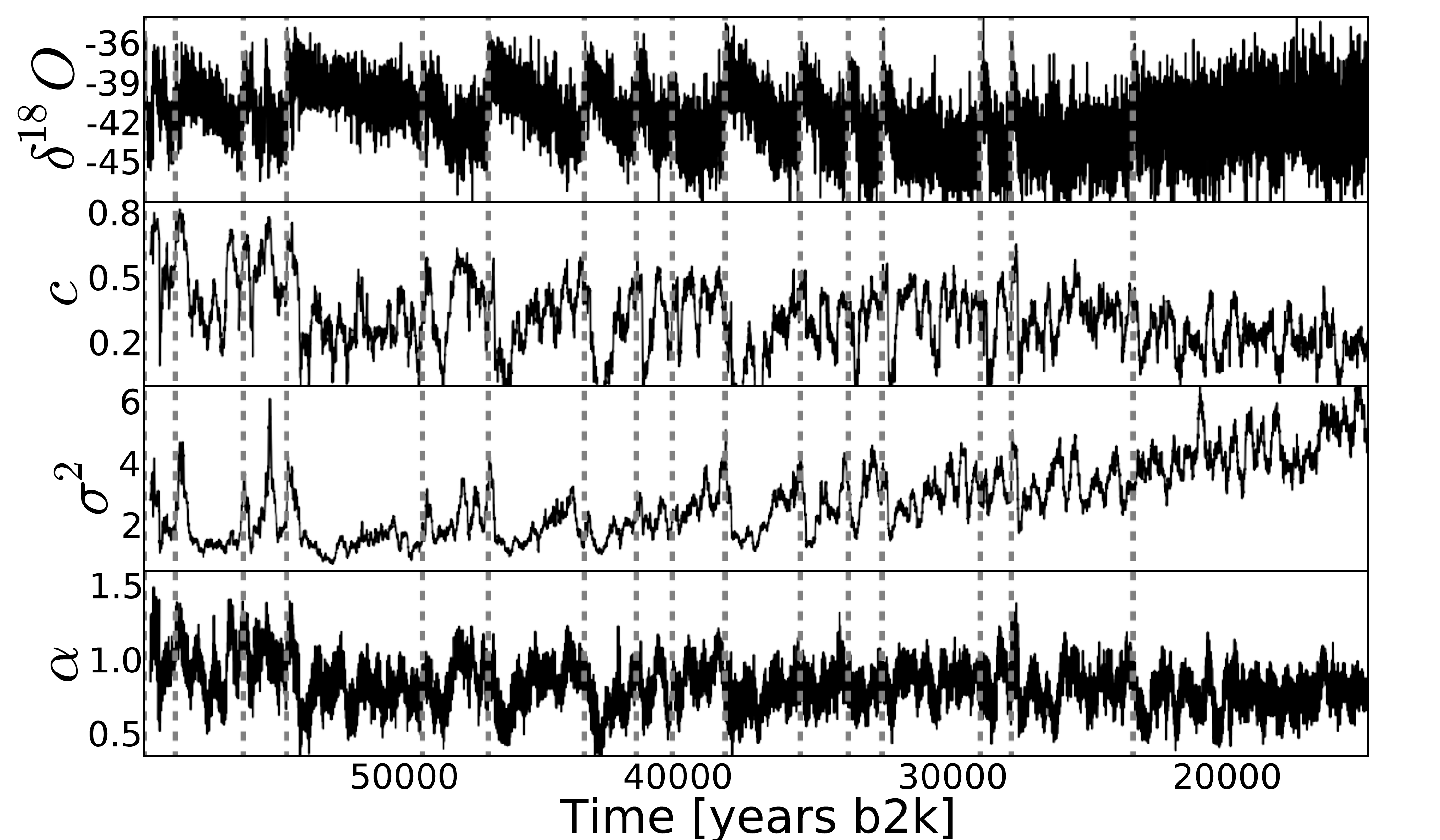}
    \caption{Ice core data analysis
    From top to bottom, the plots show the original data, the lag--1 correlation coefficient, the variance and the linear DFA exponent.
    The time scale follows the convention of years before 2,000 A.D\@. 
    Dashed grey lines mark the DO onset dating as given in~\citet{Svensson2008}.
    }
    \label{fig:TimeSeries}
  \end{center}
\end{figure}

The $\delta^{18}O$ isotope data possess a distinct trend that marks the drift of the glacial climate towards the last glacial maximum (approximately 22,000 years ago), see Figure~\ref{fig:TimeSeries} (top).
Also, associated with the drift is a gradual increase in variance that makes up an early warning signal (EWS) for the abrupt change towards the last glacial termination \citep[approximately 18,000 years ago,][]{Dakos2008,Scheffer2009}.
It turns out that the trend enhances the bimodality of the data but detrending is required to unequivocally attribute this bimodality to the DOs, as the trend is connected with a longer time scale process.

\begin{figure}[t]
  \begin{center}
    \includegraphics[width=0.5\linewidth]{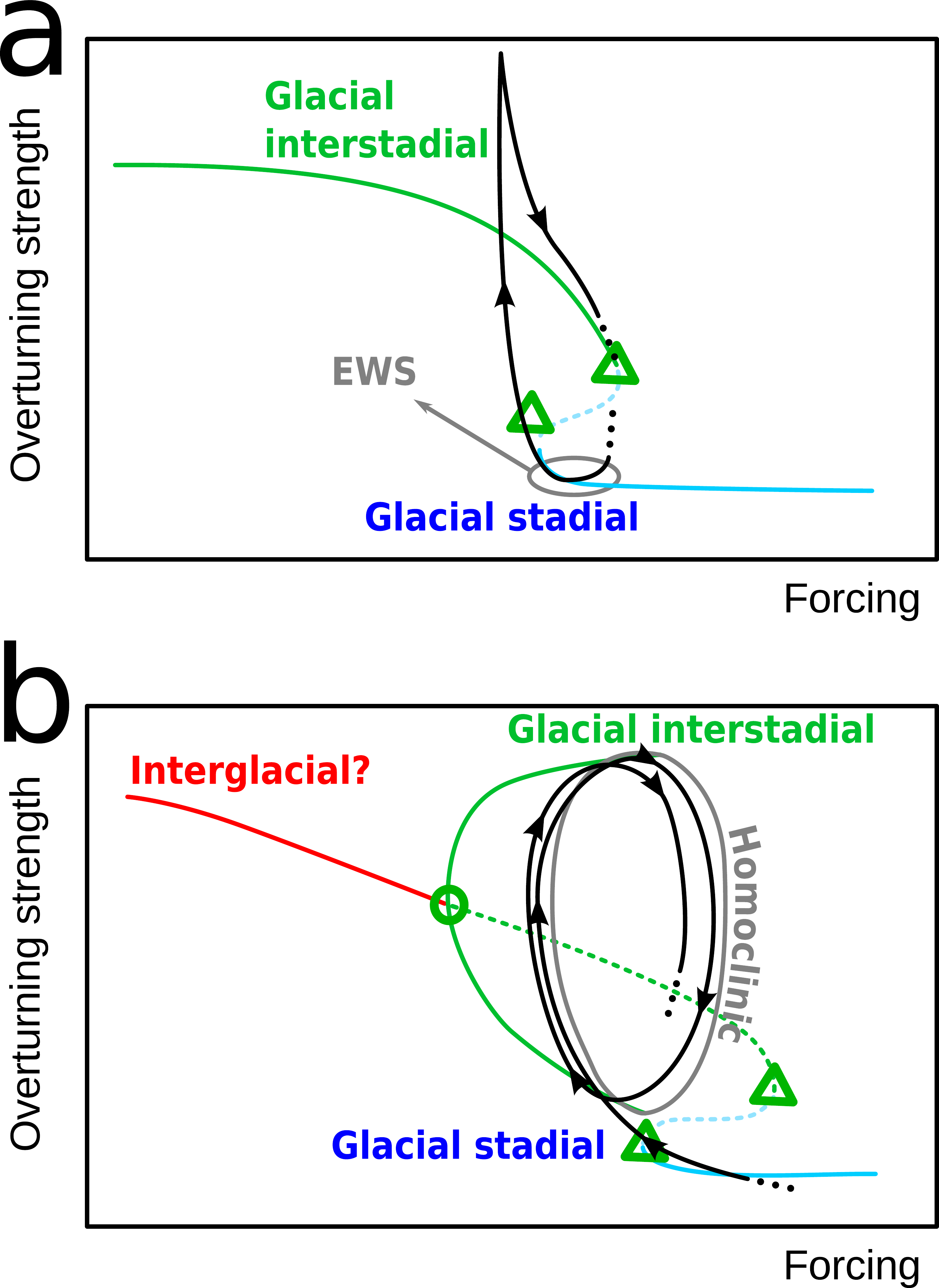}
    \caption{Sketch of prototype DO models: bifurcation points and homoclinic orbit.
    \textbf{a} Bifurcation points hypothesis. The system crosses the two bifurcation points (green triangles) periodically, due to an external forcing.
    The DOs give EWSs before the abrupt transition occurs.
    Overshooting is possible when overturning circulation recovers.
    \textbf{b} Homoclinic orbit model~\citet{Timmermann2003}. Here, DOs are due to the motion of the climate system close to a homoclinic orbit (grey) that connects to a periodic oscillation (Hopf bifurcation)~\citet{Timmermann2003,Abshagen2004,Crommelin2004} (green circle). Bifurcation points are present in the climate system, but they do not determine the abrupt transitions.
    The climate is most of the time in stadial conditions.
    A possible path followed during the oscillation is shown in black for each case.
    }
    \label{fig:sketches}
  \end{center}
\end{figure}

In the four left panels of Fig.~\ref{fig:Bimodality} (between the beginning of the time series, year 59,000b2k, and year 22,000 b2k), the normal distribution clearly lies outside the error bars for all four dimensions, and has to be rejected.
The bimodality of the data is especially evident from the PDFs of the angular coordinates.
The radial distribution is well below the normal one only at intermediate distances from the baricenter ($\rho^2\approx 5$), while it seems to be systematically above the normal distribution close to zero and in the tail, but the error bars include the normal distribution here.
These features are very robust, and they can be unambiguously detected also in phase space reconstructions using different interpolation steps, time lags and number of dimensions.
In the four right panels (after year 22,000 b2k until the end of the time series, year 15,000b2k), the deviations from normality are much smaller, and become indistinguishable from a normal distribution when the errors are taken into account.
The only statistically non insignificant deviations from normality after year 22,000 b2k are seen in panel h, which represents the longitude coordinate of the polar system.
However, no bimodality is observed, but only a non spherically--symmetrical distribution of the data.
On this basis, we can confirm the claim of \citet{Livina2011,Livina2010}; around year 22,000 b2k the climate undergoes a transition from a bimodal to a unimodal behaviour.
The climate system remains unimodal until the end of the time series (15,000 b2k); the behaviour after this time can not be assessed using this time series.
The cause of the transition at year 22,000 b2k may be a shift of the bifurcation diagram of the system towards a more stable regime (see Fig.~\ref{fig:sketches}a), either externally forced or due to internal dynamics, but may as well be caused by a reduction of the strength of stochastic perturbations.

\section{External forcing, noise or autonomous dynamics?}
\label{part:bifurcation}

In several recent works \citep{Held2004,Livina2007,Scheffer2009,Ditlevsen2010}, an abrupt transition is treated as the jump between different steady states in a system that can be decomposed in a low dimensional non--linear component and a small additive white noise component.
The jump is considered to be due to the approach to a bifurcation point, usually a fold bifurcation \citep[see e.g.][for a discussion of fold bifurcations]{Guckenheimer1988}, in which a steady state loses its stability in response to a change in the boundary conditions.
If a system perturbed by a small amount of additive white noise approaches a fold bifurcation point, an increase in the variance ($\sigma^2$) is observed, as well as in its lag--1 autocorrelation \citep{Kuehn2011,Ditlevsen2010,Scheffer2009,Held2004}.
In addition, Detrended Fluctuation Analysis (DFA) can be used to detect the approach to a bifurcation point \citep{Livina2007}.
This tool, equivalent to others that measure the decay rate of fluctuations in a nonlinear system \citep{Heneghan2000,Peng1994}, has the clear advantage over autocorrelation and variance of inherently distinguishing between the actual fluctuations of the system and trends that may be part of the signal.
This follows from the fact that the detrending is part of the DFA procedure itself, and different orders of detrending can be tested, provided that a sufficient amount of data is available to sustain the higher order fit.
Approaching a bifurcation point, the lag--1 autocorrelation coefficient ($c$) and the DFA exponent ($\alpha$) both tend to increase, ideally towards $1$ and $3/2$ respectively \citep{Held2004,Livina2007}.
This reflects the fact that the fluctuations in the system are increasingly correlated as the bifurcation point is approached: they become more similar to a Brownian motion as the restoring force of the system approaches zero.
It must be reminded that all the results obtained using these techniques are based on strong assumptions on the nature of the noise.
The noise in the signal is generally assumed to be of additive nature, to be colorless and to be recorded in the time series through the same processes that give rise to the slower dynamics.
A clear separation of time scales between the noise and the low--dimensional dynamics is also assumed.
These assumptions may not be satisfied by the climate system, and this important caveat has to be considered when evaluating the results.

Our aim is to analyse the statistical properties of the shorter time--scale fluctuations in the ice core record that occur before the onset of DOs.
The quality of the paleoclimatic time series and the intrinsic limits of the EWS technique are a serious constraint to the conclusions that can be drawn.
Still, the analysis identifies EWSs before the onset of DOs.
This suggests that the onset of DOs is connected with the approach to a point of reduced stability in the system.
This finding is compatible only with some of the models proposed as an explanation of DOs.
Such a selection is possible since the different prototype models have different, peculiar, ``noise signatures'' before an abrupt transition takes place.
It must be stressed that the problem is an inverse one; multiple mathematical models qualitatively consistent with the ``noise signatures'' can be developed, and only physical mechanisms can guide the final choice.

We will focus on the three characteristics of the data mentioned above: the correlation coefficient of the time series at lag--1, the variance, and the DFA exponent.
They are plotted, together with paleoclimatic time series used, in fig.~\ref{fig:TimeSeries}.
A similar type of analysis has been used before in the search of EWSs for abrupt climate transitions \citep[see e.g.][]{Held2004,Livina2007,Scheffer2009,Ditlevsen2010}.
We use here EWSs to put observational constraints to the various prototype models for DOs, considering only the properties of the ensemble of events.
In contrast to earlier studies we consider an EWS not as a precursor of a single event, but analyse instead the average characteristics of the EWS found in the ensemble of DOs.
It must be stressed that \citet{Ditlevsen2010} did consider the ensemble of events, but did not compute any average quantity.
Given the low to signal--to--noise ratio, computing the mean of the ensemble and an error estimate of it may uncover signals otherwise too weak to be detected.
This approach is more robust than the one which considers each realisation (each DO) separately, since EWSs have a predictable behaviour only when computed as an average over an ensemble of events, as demonstrated by \citet{Kuehn2011}.
We argue that only considering the mean of the events we can hope to draw general conclusions on the nature of these climate signals.

\subsection{Prototype models and early warning signals}

The reason why we can discriminate between various prototype models is that, in the case of a system that crosses, or comes close to a bifurcation point, $\sigma^2$, $c$ and $\alpha$ all increase before abrupt change occurs, while for other prototype models this is not the case.

For each prototype model that we considered, the relevant equation, or set of equations, is integrated with an Euler--Maruyama method~\citep{Higham2001}.
The associated time series is then analysed with respect to lag--1 autocorrelation, variance, and the DFA exponent (details of the analyses are given in section~\ref{sec:EWSmethods}).

\subsubsection{Bifurcation point in a double well}
\label{sec:bifurcation}

EWSs are visible if the DOs are caused by a slowly changing external forcing, in which a critical parameter, like northward freshwater transport, varies over a large enough range for the meridional overturning circulation of the ocean to cross the two bifurcation points that mark the regime of multiple equilibria.
This idea, in the context of paleoclimate, dates back to \citet{Paillard1994} who suggested it for explaining Heinrich events, and is often invoked more in general as a paradigm for understanding the meridional overturning circulation stability \citep[see e.g.][]{Rahmstorf2005}.
In physical terms, this model implies that a changing external forcing is causing a previously stable state to disappear.
In Figure~\ref{fig:sketches}a this situation is sketched.

\begin{figure}[t]
  \begin{center}
    \includegraphics[width=1.\textwidth]{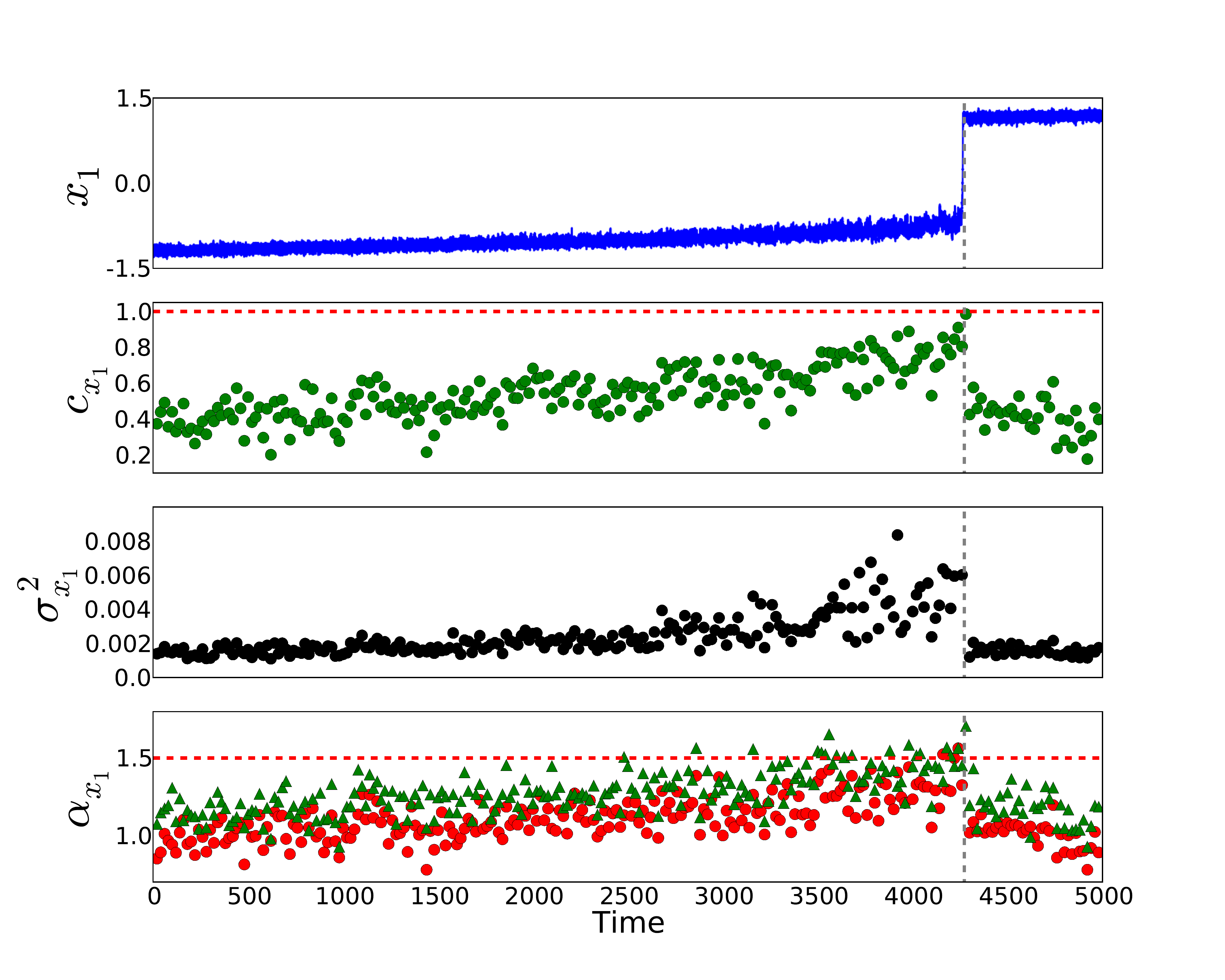}
    \caption{Bifurcation point mechanism: double well potential under changing external forcing.
    An external parameter is slowly changed forcing the system to undergo an abrupt transition, that shows clear EWSs.
    The top panel (blue line) shows the time series for this system.
    Below, the lag--1 autocorrelation ($c$, green circles) and variance ($\sigma^2$, black circles) are shown.
    In the bottom panel, DFA exponent $\alpha$ for linearly (red circles) or quadratic (green triangles) detrending case.
    Grey vertical line marks the abrupt transition onset, the red line marks a critical $c$ or $\alpha$ value.
    \label{fig:DoubleWell}
    }
  \end{center}
\end{figure}

As a prototype for a system crossing a bifurcation point, the same equation used in~\cite{Ditlevsen2010} is used:
\begin{equation}
\dot{x} = -x^3 + x + q + \sigma \eta(t), \label{eq:doubleWell}
\end{equation}
where $x$ is the state variable (time dependent), $\dot{x}$ is its time derivative, $q$ is the only system parameter and $\sigma \eta(t)$ represents a white noise (normal distribution), with standard deviation given by $\sigma$.
In this case, the evolution of the system can be thought of as the motion of a sphere on a potential surface that has either one or two stable equilibrium states, divided by a potential ridge.
This potential surface can change in time, due to slow changes in the parameter $q$.
In this case, $q$ will be time dependent, and will represent a forcing external to the dynamics of the system, represented only by the scalar function $x$.
When $q$ crosses one of the two bifurcation points ($q_0=\pm 2\sqrt{3}/9$), the system changes from having two possible equilibrium states to having only one.
At these points, one of the two states loses its stability and the system can jump to the second equilibrium (Figure~\ref{fig:DoubleWell}, shown with $\sigma=0.1$ and $q$ going from $-0.5$ to $0.5$ during the integration).

\subsubsection{Noise induced transition}

\begin{figure}[t]
  \begin{center}
    \includegraphics[width=1.\textwidth]{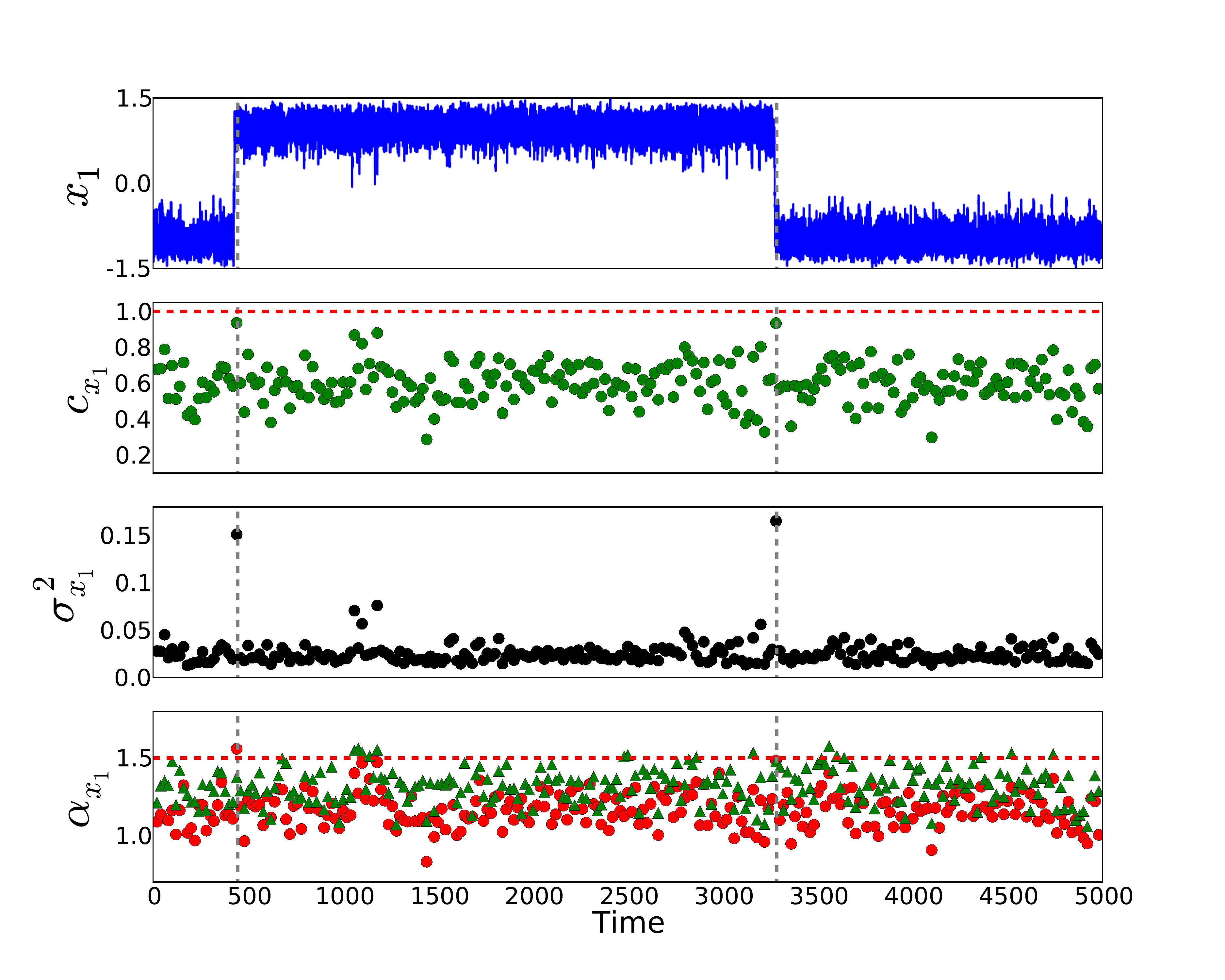}
    \caption{As in Figure~\ref{fig:DoubleWell}, but for a noise induced transition.
    The external forcing is kept constant, but the noise can still trigger abrupt transitions between two available states.
    No trend in the statistical properties computed is detected, thus no EWS is observed.
    \label{fig:Noise}
    }
  \end{center}
\end{figure}

If $q$ is instead constant in time (and the system is thus an autonomous one), abrupt transitions can still occur if more than one equilibrium state is available, and noise is sufficiently strong to trigger the transition.
An example of this case, in the simple case of a fixed double well with constant white noise, is shown in Fig.~\ref{fig:Noise} (in this case, $\sigma=0.3$ and $q=0$).
In this model, the system is normally in the cold stadial state but can jump due to the noise in the forcing, say atmospheric weather, to a warmer, interstadial state.
In this case no trend is visible in the statistical properties of the time series, thus no EWS can be detected.
The two cases described are those considered by \citet{Ditlevsen2010}, to which we refer for a more detailed discussion.

\subsubsection{Stochastic resonance}
\label{sec:stochasticRes}

\begin{figure}[t]
  \begin{center}
    \includegraphics[width=1.\textwidth]{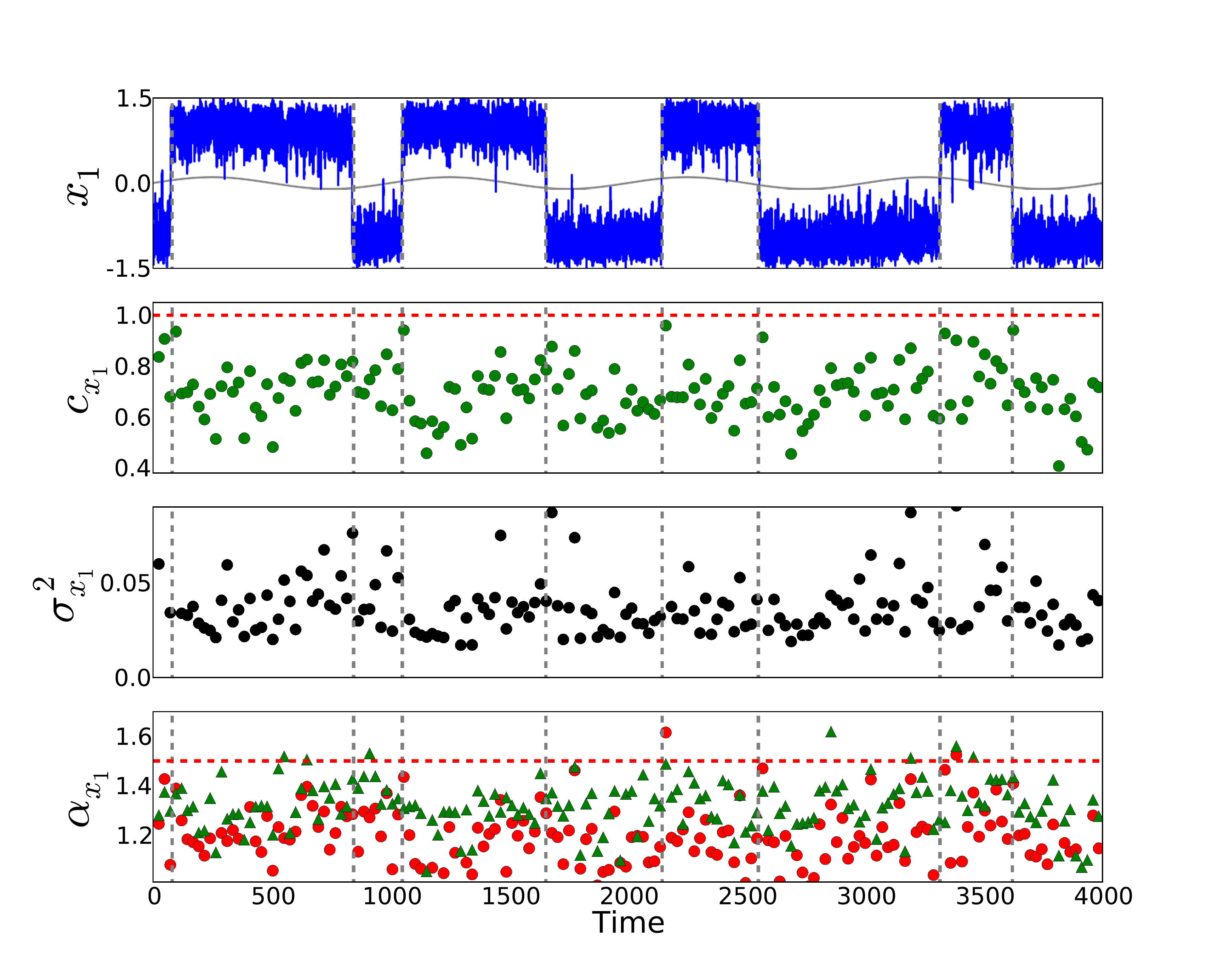}
    \caption{Stochastic resonance.
    Same as in Figure~\ref{fig:DoubleWell}, but for the case of stochastic resonance.
    The forcing oscillates, and transitions take place preferentially when the system is closer to the bifurcation point.
    The deterministic part of the forcing (see text) is shown as a gray line on the top panel.
    \label{fig:StocRes}
    }
  \end{center}
\end{figure}

A slightly different case, again based on the model of Eq.~\ref{eq:doubleWell}, is that discussed by \citet{Alley2001,Ganopolski2001,Ganopolski2002}.
They suggest that DOs may be connected to stochastic resonance, where the parameter $q$ in Eq.~\ref{eq:doubleWell} becomes a periodic function of time.
In this case, transitions from one state to the other take place preferentially when $q$ is closer to the bifurcation point, but are still forced by the noise term rather than by $q$ actually reaching its bifurcation point.
In Fig.~\ref{fig:StocRes} this case is explored, using $q\equiv q(t)=q_0 \mathrm{sin}(\frac{2 \pi t}{\tau})$ with $q_0=0.1$ and $\tau=1000$, the noise level is $\sigma=0.35$.
In this case, EWSs are clearly present before some of the events (e.g.~the second and the fourth), while for others no clear EWS is detected.\footnote{The upwards and downwards transitions are, on average, equivalent, similarly to the case of noise induced transitions with $q=0$.}
EWSs are present in some cases, while absent in others, since transitions take place preferably when $q$ is closer to the bifuration point, but not only in this phase.
Transitions can still take place purely due to the presence of noise, and in those cases no clear EWS will be detected.
If the average from a sufficiently large ensemble of events is considered EWSs can be detected unambiguously.
A very similar situation will be found in the paleoclimatic data, and the presence of EWSs will be shown there in the ensemble average.

\subsubsection{Other models}

We will show that other models proposed in the literature for explaining DOs do not posses any EWS.
No EWS can be detected if DOs are due to an unforced (ocean) mode of oscillation \citep{Sakai1999,Verdiere2006} since, for an autonomous oscillation, there are in general no changes in the stability properties of the system while it evolves in time \citep[see e.g.~][]{Guckenheimer1983}.
DOs have been linked to oscillations close to a ``homoclinic orbit'' by \citet{Timmermann2003}.
A homoclinic orbit is a periodic solution to a system of ordinary differential equations that has infinite period, but in the presence of a small amount of noise it features finite return times.
A sketch of this prototype model is shown in Figure~\ref{fig:sketches}b.
To describe this model the set of Ordinary Differential Equations given in~\citet{Crommelin2004} is used.
This minimal mathematical model has been suggested before in the context of atmosphere regime behaviour \citet{Charney1979}, but it is mathematically equivalent to the mechanism for DOs suggested in~\citet{Timmermann2003,Abshagen2004}.

\begin{figure}[t]
  \begin{center}
    \includegraphics[width=1.\textwidth]{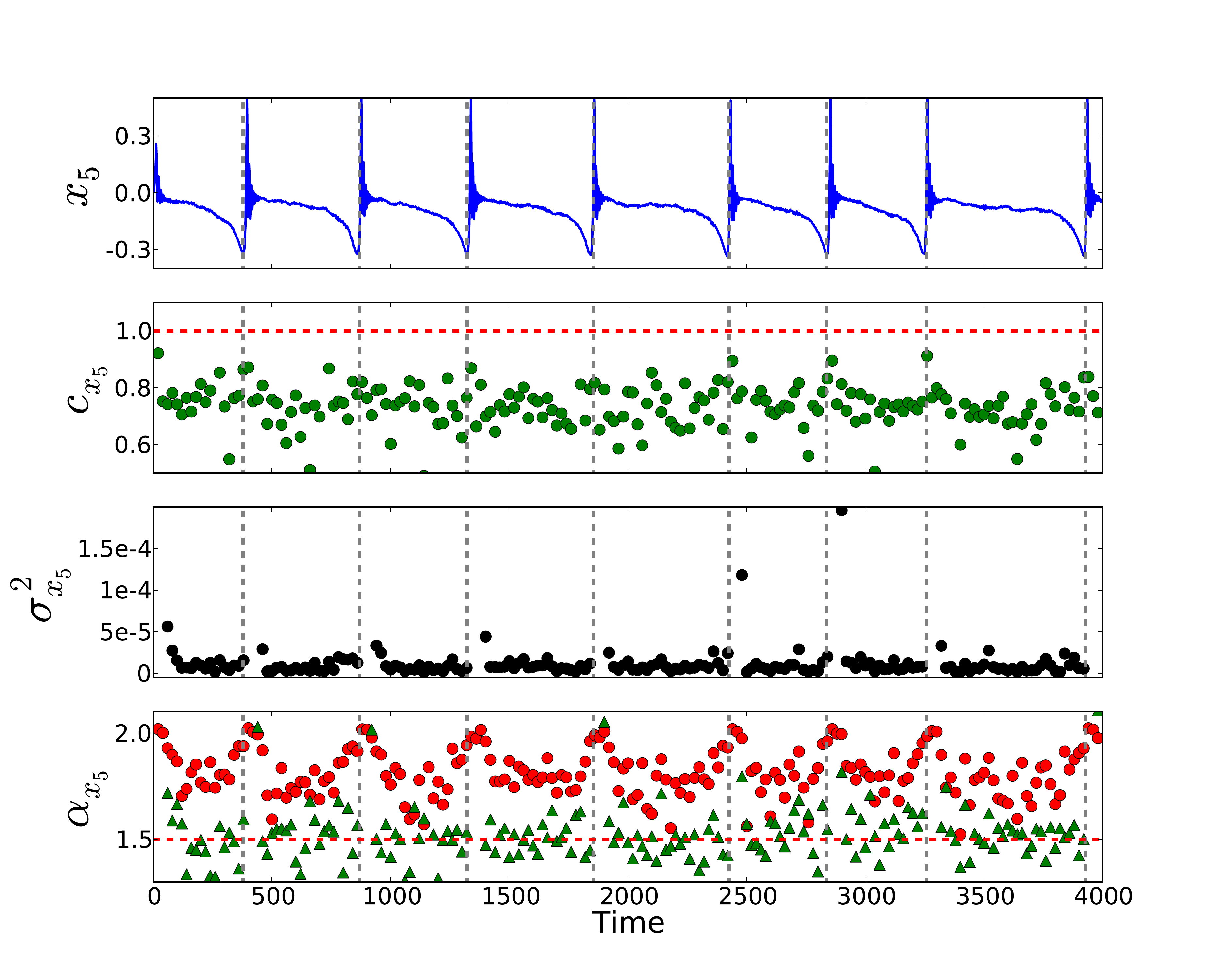}
    \caption{Charney and Devore model.
    As in Figure~\ref{fig:DoubleWell}, for the Charney and DeVore model.
    In this case, the abrupt transitions are not connected with a changing external forcing, but are an autonomous mode of the equations.
    Dimension number 5 ($x_5$) of the 6--dimensional ODE system is considered.
    It is evident that, for $\alpha$ (bottom panel), a detrending of order higher than one is needed to remove the effect of non--stationarities in the time series.
    After proper detrending, no EWSs are observed in $\alpha$.
    Red circles (green triangles) refer to the linear (quadratic) DFA.
    \label{fig:CharneyDevore}
    }
  \end{center}
\end{figure}

To investigate EWSs for a periodic cycle close to a homoclinic orbit, the following system of ordinary differential equations \citep{Crommelin2004} is integrated (see Figure~\ref{fig:CharneyDevore}):
\[
\begin{split}
\dot{x_1} &= \widetilde{\gamma}_1 x_3 - C (x_1 - x_1^*) \\
\dot{x_2} &= -(\alpha_1 x_1 - \beta_1) x_3 - C x_2 - \delta_1 x_4 x_6\\
\dot{x_3} &= (\alpha_1 x_1 - \beta_1) x_2 - \gamma_1 x_1 - C x_3 + \delta_1 x_4 x_5\\
\dot{x_4} &= \widetilde{\gamma}_2 x_6 - C (x_4 - x_4^*) + \varepsilon (x_2 x_6 - x_3 x_5)\\
\dot{x_5} &= -(\alpha_2 x_1 - \beta_2) x_6 - C x_5 - \delta_2 x_3 x_4\\
\dot{x_6} &= (\alpha_2 x_1 - \beta_2) x_5 - \gamma_2 x_4 - C x_6 + \delta_2 x_2 x_4.\\
\end{split}
\]
The system has six dimensions ($x_i$) with parameters as in~\cite{Crommelin2004}, chosen as they produce the abrupt, quasi periodic transitions observed in Fig.~\ref{fig:CharneyDevore}, with $\widetilde{\gamma}_1=0.06$, $C=0.1$, $x_1^* = 0.89$, $\alpha_1=0.24$, $\beta_1=0.25$, $\delta_1=0.384$, $\gamma_1=0.048$, $\widetilde{\gamma}_2=0.024$, $x_4^*=-0.82325$, $\varepsilon=1.44$, $\alpha_2=0.734$, $\beta_2=0.0735$, $\delta_2=-1.243$.
White noise is added to each component, the noise has standard deviation of $0.001$.
This model is chosen as it provides a simple description of the results of~\cite{Timmermann2003}, having a very similar bifurcation diagram (the fundamental element is the presence of a fold--Hopf bifurcation).
As the time series is strongly non--stationary, linear DFA is not a good measure of the fluctuation decay in this case, and higher order detrending is needed (orders higher than 2 give results consistent with the case of quadratic detrending).

After integrating these equations, again no precursor for abrupt changes can be detected (Fig.~\ref{fig:CharneyDevore}).
It must be stressed that, in order to correctly compute the statistical properties of the time series, it is of fundamental importance that the data are detrended (linear or higher order detrending), to ensure that the findings can be ascribed to random fluctuations and not to the non--stationarity of the data.
In particular, in the case of the DFA exponent computation for the latter prototype model, the strong non--stationarity of the data requires a quadratic detrending.

The results from other models that have been suggested as a prototype for oscillations close to a homoclinic orbit~\citep{Welander1982,Stone1999} consistently show no EWS before abrupt transitions (not shown).

\subsection{Fluctuations analysis}
\label{sec:EWSmethods}

We now have seen that EWSs are characteristic of abrupt changes connected with the approach to a bifurcation point.
The bifurcation point is either crossed while a parameter of the system is changed (Sec.~\ref{sec:bifurcation}) or is only approached, without actually crossing it, as in the case of stochasitc resonance (Sec.~\ref{sec:stochasticRes}).
If the ice core record also features EWSs, the other prototype models are falsified on these bases.
The ice core data are shown in Figure~\ref{fig:TimeSeries}.
The EWSs of the time series are computed in a sliding window 250 years wide, and the results are plotted at the right end of the sliding window to avoid contamination from points after the DO onset.
This window size corresponds on average to 100 data points for the variance computation.
For $c$ and $\alpha$, the time series is first linearly interpolated to a step of 4 years and approximately 60 points are used in the window.
The time step is chosen in order to avoid overfitting the data even in the parts of the time series with lower resolution.
The window width choice is a compromise between the need to minimise noise (larger window) and to maximise the number of independent points for the computation of trends (smaller window).
The data is linearly detrended in the window before the computation, to compensate for potential nonstationarities in the time series.
In the computation of the DFA exponent \citep{Peng1994}, 10 box lengths were used, ranging from 16 to 52 years.
The number of time lengths is limited for computational reasons.
Different time steps for the interpolation and boxes lengths were tested.
The results show small sensitivity to these choices, as long as the time step for the interpolation is kept below approximately 10 years.

If one considers the successive DOs one by one, it is clear that for some of the DOs a marked increase is found (e.g.~the one at approximately 45,000 years b2k), implying an EWS, but for other DOs no EWSs can be detected (e.g.~the last one).
This has to be expected in general from a stochastic system for which only the average behaviour is predictable ~\citep{Kuehn2011}, and may also be the fingerprint of stochastic resonance, as discussed in Sec.~\ref{sec:stochasticRes}.

\subsection{Results}

\begin{figure}[t]
  \begin{center}
    \includegraphics[width=0.5\linewidth]{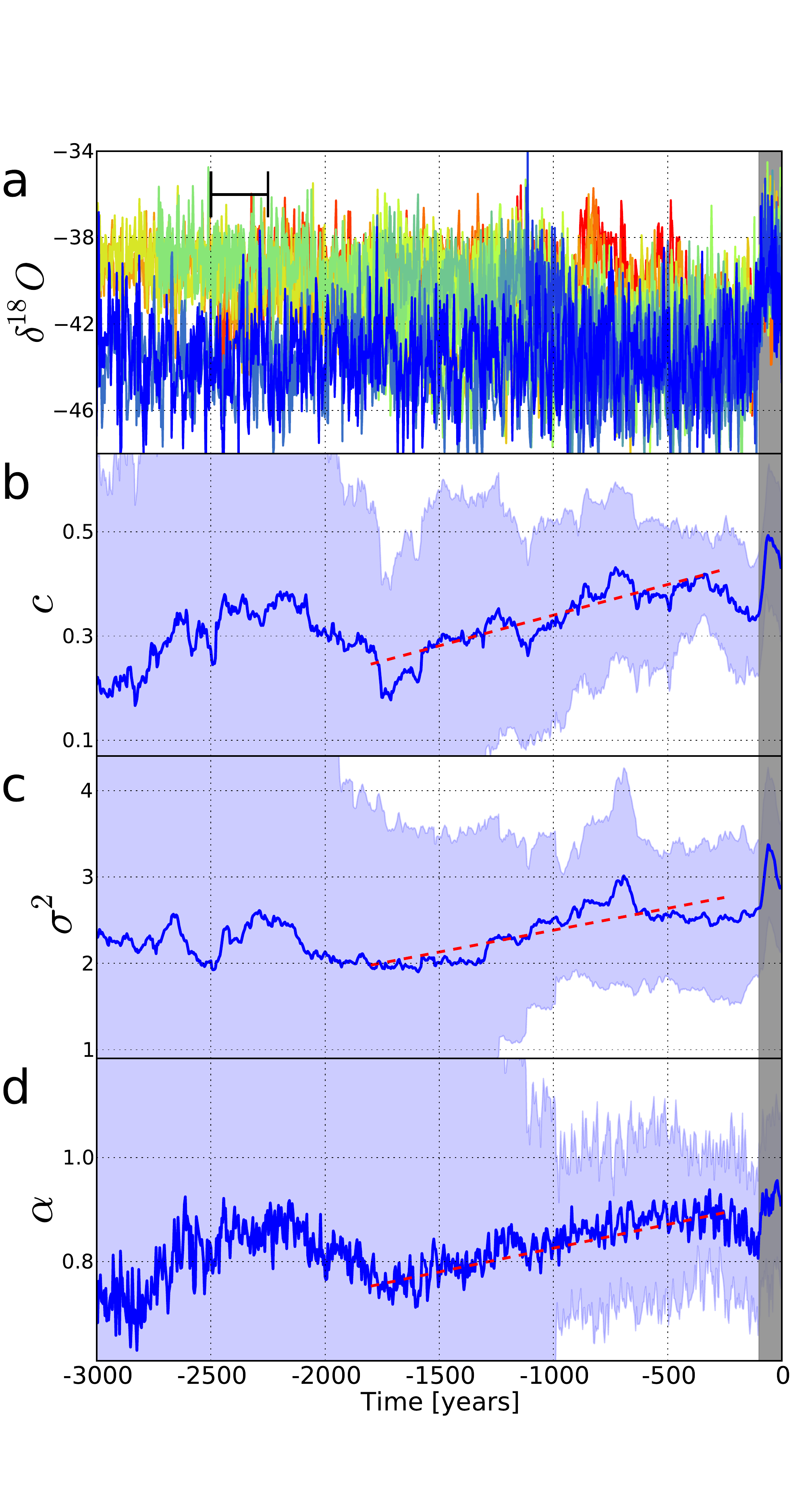}
    \caption{Ensemble analysis.
    \textbf{a} Time series of DOs number 2 to 16 from the data of \citet{NGRIP2004,Svensson2008}.
    The time series are aligned so that the DOs start at year -100 (grey shading after event onset). 
    Colours mark different events, from number 2 (blue) to number 16 (red).
    The time window used in the computations is drawn in black.
    Below, the lag--1 autocorrelation (\textbf{b}), variance (\textbf{c}) and linear detrended fluctuation analysis exponent (\textbf{d}).
    The blue line is the ensemble average, the shaded area marks the standard deviation.
    The dashed red line shows a least square regression of the data over the marked time range.
    \label{fig:DOensemble}
    }
  \end{center}
\end{figure}

As discussed by \citet{Kuehn2011}, analysing the properties of an ensemble of events, instead of a single realisation, is a more robust approach and better justified on theoretical grounds.
We thus use EWSs to characterise the properties of the ``mean DO'' instead of trying to predict the onset of the following transition.
For this reason, we consider the whole ensemble of DOs, instead of each single DO.
With this aim, the time series is cut into slices that end 100 years after the transition onset and start either 100 years after the previous DO onset or, if the time span between two DOs is sufficiently long, 2,900 years before the following DO onset (Figure~\ref{fig:DOensemble}a).
The onset of the transitions then are translated to year -100.
In this way, an ensemble of DOs is defined, being composed of slices of the original time series of variable length, synchronised at the onset of each DO.
If the quantities used for EWS detection are then averaged for the whole ensemble, a moderate but significant increase is observed in all three fluctuation properties, starting approximately from year -1,800 until year -250 (Figure~\ref{fig:DOensemble}b--d).
The standard deviation of the ensemble is large, in particular at times far from the DO onset because the ensemble has fewer members there (at times smaller than -1000 years), but the trend can not be discarded as a random fluctuation for all three cases.
To test this, a linear least square fit of the data in the interval -1,800 to -250 is performed, only using independent points from the ensemble (thus at a distance of 250 years from each other, given the window width of 250 years), and obtaining an error interval from a bootstrapped ensemble of 50,000 members.
The results of this fitting procedure are reported in Table~\ref{tab:LinFit}.
In all three cases the linear trends are well above their standard deviation, providing a strong indication of the robustness of the signal.
In order to check the robustness of the findings on the order of detrending for DFA, the computation has been repeated with a quadratic detrending, actually giving a slightly stronger trend (see Table~\ref{tab:LinFit}).
These results are consistent with a scenario that connects DO onset with either the crossing or the approach of a bifurcation point.
In other words, these findings are consistent with either a model where the system is forced to shift from a steady state to a different one, or with the stochastic resonance mechanism, where transitions take place preferentially closer to the bifurcation point, even if this one is not actually reached, and transitions are due to the noise.

The fact that $c$ and $\alpha$ do not reach their theoretical values for a bifurcation point, respectively $1$ and $3/2$, can easily be explained: the noise in the system triggers the transition before the actual bifurcation point is reached.
This clearly must be the case for stochastic resonance, but is relevant also if the bifurcation point is crossed \citep{Meunier1988}.
Also, the three quantities $c$, $\sigma^2$ and $\alpha$ decrease again in the last few hundred years before the DO onset.
This may be a bias in the relatively small ensemble considered here, but a similar decrease has been shown by \citet{Kuehn2011} for various bifurcations in fast--slow systems.
For several idealised systems containing different forms of bifurcation points (included the fold bifurcation considered here), a decrease in variance in the immediate vicinity of the bifurcation point is found, while the well known increase is observed when the bifurcation point is farther away in the parameter space.
To confirm that the decrease observed in the data is consistent with the one discussed by \citet{Kuehn2011}, the time scale of the decrease should be linked to the distance of the bifurcation parameter from the bifurcation point.
Unfortunately, this is not possible, as we have no information on the parameter range that may be spanned by the real system.
Variance in particular starts to decrease quite far from the DO onset (approximately 700 years).
This may indicate a weakness of the results (even if, as discussed above, the increase is still statistically significant), but it has been shown that variance may not be the most robust indicator of EWS, and autocorrelation may be given more emphasis \citep[see][]{Dakos2012}.
The observation of a decrease in variance just before the DO onset, after a longer period of increase, is an important lesson for the practical application of these techniques in the detection of approaching bifurcation points.

Our findings are in contrast with the previous suggestions of \citet{Ditlevsen2010}.
\citet{Ditlevsen2010} considered all the events together, without considering neither the ensemble mean nor the individual events.
This approach may prevent from uncovering a weak mean signal, as well as the clear EWS visible before some of the events.
We have seen in particular that in the case of stochastic resonance the presence of EWS is not guaranteed before each abrupt transition.
Furthermore, strong noise may hide the increase in the indicators for some of the events even if a bifurcation is actually crossed \citep{Kuehn2011}.
We do not think that our different results may be ascribed to a different choice in the parameters of the analysis, as several different parameter sets have been used, and the results are robust.

Still, an important caveat must be reminded, relevant for most of the works dealing with EWSs.
The signal detected as an EWS is described by the simple models discussed, but other models may give a similar signal as well; here we try to distinguish only among the models that have been proposed to describe DOs.
Furthermore, the effect of other types of noise other than white and additive are not studied.
This is the most common approach in the field \citep[one of the few exceptions, to our knowledge, is][]{Bathiany2012}.
A closer investigation of the effects of using red and multiplicative noise is an interesting topic, but outstide the scope of this paper.

From a broader perspective, the motivation for classifying DOs into three groups\footnote{Stochastic resonance is a fourth group, which can be considered as a hybrid between noise induced transitions and bifurcation induced transistions.} may seem unclear, but this is instead a very important point.
If DOs are associated with the presence (or the crossing) of a bifurcation point, as our results suggests, this would mean that the climate can potentially show hystheresis behaviour, i.e.~abrupt transitions in opposite directions take place under different forcing conditions, and abrupt transitions are thus to some extent irreversible, with obvious implications e.g. for global warming.

\section{Summary}

In this work, we performed two sets of analysis on a well-known paleoclimatic record, the $\delta^{18}O$ isotope data from the NGRIP ice core \citep{NGRIP2004,Svensson2008}, believed to be representative of temperature over the North Atlantic sector.
We assessed bimodality of the system using a phase--space embedding technique, that guarantees that the bimodality is not an artifact of projection of the complex dynamics of the climate system on a scalar time series.
We confirm with this technique the claim of \citet{Livina2011,Livina2010}, that a switch from bimodal to unimodal behaviour is observed in the time series around year 22,000 before present.

Secondly, we analysed the statistical properties of the fluctuations in the paleoclimatic time series, before the onset of DOs.
In particular, we focused on the average properties of the events considered as an ensemble instead of each separately.
Despite the high level of noise in the data, EWSs can be detected in the ensemble average, consistently with the hypothesis that DOs take place preferentially close to a bifurcation point in the climate system.
In particular, our findings seem to be particularly close to the stochastic resonance scenario proposed by \citet{Alley2001,Ganopolski2001,Ganopolski2002}.
Other prototype models that have been proposed \citep{Timmermann2003,Sakai1999,Ditlevsen2010} are less consistent with the data, as their mechanisms do not involve any transition from which EWSs can, at least in general, be expected.
\citet{Ditlevsen2010} came to opposite conclusions, but they did not consider the average behaviour of the ensemble, while we think that this may be a step of fundamental importance.

A disclaimer has to be made: our conclusions hold for the ensemble of the events, and are a probabilistic statement: we can only claim that a scenario that does not include an approach to, or the crossing of, a bifurcation point with EWS is unlikely.
The trends of Table~\ref{tab:LinFit} are between two and three standard deviations apart from zero.
This means that the probability that the real trend in each indicator is zero is low but non--zero, in the order of $1\%$, assuming a normal distribution.
Given the rather complex techniques used, we can not rule out the possibility that the error estimates given in the Table may be underestimated.
Apart from the general limitations of the techniques used, we also want to remind that we considered only models already discussed in the literature in the context of DOs.
Other models, giving similar signals, may be developed, but given the inverse nature of the problem faced here, the models considered must be limited in number.

A connection with the meridional overturning circulation instability remains in the domain of speculation, but is a plausible hypothesis considering the large evidence linking DOs signals to rapid changes in the meridional overturning circulation.
Further investigation is needed to confirm this hypothesis and, more importantly, to address the fundamental question that remains open: if DOs are due to an external forcing, what is the nature of this forcing?
Furthermore, the relation between the variability recorded in the $\delta^{18}O$ time series and the AMOC variability is still uncertain, and the link between the two may be far from direct, involving e.g.~atmospheric or sea ice variability.

Looking beyond the results of the particular time--series used here, we suggest that EWSs may provide a useful guide for discriminating between different models meant to describe a climate signal.
When data is scarce, the analysis of the average properties of the fluctuations can give important hints on the nature of the signal which produced it.

\begin{acknowledgments}
The authors acknowledge Peter Ditlevsen (University of Copenhagen) for discussion and providing the data, Anders Svensson (University of Copenhagen) for making the data available, Henk Dijkstra (University of Utrecht) for suggesting to use Charney--Devore model as a prototype for homoclinic orbits and for pointing out some inconsistencies in an earlier version of the manuscript, Timothy Lenton (University of Exeter) for valuable comments.
A.A.C. acknowledges the Netherlands Organization for Scientific Research for funding in the ALW program.
V.L. acknowledges NERC and AXA Research Fund for funding.
The authors would also like to thank the reviewers for their comments and precious suggestions.
\end{acknowledgments}

\clearpage

\begin{table}[t]
  \begin{center}
    \begin{tabular}{c || c | c}
      Quantity & Linear trend & Standard deviation\\
      \hline
      $c$ & $1.2\cdot 10^{-4}$ & $3 \cdot 10^{-5}$ \\
      $\sigma^2$ & $5.1\cdot 10^{-4}$ & $2 \cdot 10^{-4}$ \\
      $\alpha$ & $9.1\cdot 10^{-5}$ & $3 \cdot 10^{-5}$ \\
      $\alpha$ (quadr.)& $1.2\cdot 10^{-4}$ & $4 \cdot 10^{-5}$ \\
    \end{tabular}
  \end{center}
  \caption{\textbf{Linear trends of EWS.}
  Results of the linear fit of the trends in the ensemble mean for autocorrelation ($c$), variance ($\sigma^2$) and DFA exponent ($\alpha$) in the time interval from -1800 to -250 years before the DO onset. 
  The last line refers to the result of the fit from the quadratic DFA (see text).
  The error values are computed from a bootstrapped ensemble of 50,000 members.
  \label{tab:LinFit}
  }
\end{table}

\end{document}